\documentclass{sadhana}

\usepackage[normalem]{ulem}
\usepackage{graphicx,amsmath,bm}
\usepackage{algorithm}
\usepackage[noend]{algpseudocode}
\usepackage{textcomp}
\usepackage{listings}
\usepackage{hyperref}
\usepackage{alltt}
\usepackage{tikz}
\usepackage{framed}
\usepackage[framemethod=TikZ]{mdframed}
\usepackage{crayola}
\usepackage{xcolor, colortbl}
\usepackage{mathpartir}
\usepackage{tabularx}
\usepackage[skins]{tcolorbox}
\usepackage{multicol}
\usetikzlibrary{automata,positioning,shapes,arrows, backgrounds, fit, shadows}
\usetikzlibrary{decorations.markings, calc}
\usetikzlibrary{arrows.meta}
\usetikzlibrary{decorations.pathmorphing}
\usepackage{mathtools}
\usepackage{array}
\usepackage{multirow}
\usepackage{amsmath}

\newtheorem{example}{Example}
\begin{document}

%%paper title
%%For line breaks, \\ can be used within title 
\title{Automated Application Processing}

%%author names are separated by comma (,) 
%%use \and before the last author name 
%%\textsuperscript{number} is used for affiliation
%%use a * along with the number separated by comma
%% for the  author for correspondence

\author{Eshita Sharma, Keshav Gupta, Lubaina Machinewala, Samaksh Dhingra, Shrey Tripathi, Shreyas V S \and Sujit Kumar Chakrabarti}
\affilOne{International Institute of Information Technology, Bangalore}

%%escape two column mode for title, affiliation and abstract
%%by giving \twocolumn command as shown

\twocolumn[{

\maketitle

%%abstract

\begin{abstract}
Recruitment in large organisations often involves interviewing a large number of candidates. The process is resource intensive and complex. Therefore, it is important to carry it out efficiently and effectively. Planning the selection process consists of several problems, each of which maps to one or the other well-known computing problem. Research that looks at each of these problems in isolation is rich and mature. However, research that takes an integrated view of the problem is not common. In this paper, we take two of the most important aspects of the application processing problem, namely \emph{review/interview panel creation} and \emph{interview scheduling}. We have implemented our approach as a prototype system and have used it to automatically plan the interview process of a real-life data set. Our system provides a distinctly better plan than the existing practice, which is predominantly manual. We have explored various algorithmic options and have customised them to solve these panel creation and interview scheduling problems. We have evaluated these design options experimentally on a real data set and have presented our observations. Our prototype and experimental process and results may be a very good starting point for a full-fledged development project for automating application processing process.

\end{abstract}

%%manuscript information
%%received, revised and accepted dates
%%
%\msinfo{1 October 2021}{1 December 2021}{1 January 2022}

%%insert keywords separated by comma

}]
%%close the twocolumn escape here

%%include \corres to print the footnote for correspondence
%%include \volnum{number} for the volume number in the header
%%include \monthyear{month year} for month and year of publication in the header
%%include \pagefirst{num} first page number of article in header
%%include \pagelast{num} last page number of article in header
%%include \doi{number} for the DOI number in the header

%\setcounter{page}{2333}
%\corres
%\volnum{123}
%\issuenum{4}
%\monthyear{January 2022}
\pgfirst{1}
%\pglast{2335}
%\doinum{12.3456/s78910-011-012-3}
\articleType{}%%use \articleType{L} for Sadhana Letters, empty/null otherwise

%%The running head information

\markboth{Sharma, Gupta, Machinewala, Dingra, Tripathi, Shreyas, Chakrabarti}{Automated Application Processing}

\section{Introduction}
For large organisations, effective talent management critical for organisational success. Not surprisingly, organisations invest heavily to ensure that the best talents are hired. It is important to carry this process smoothly/efficiently, otherwise, the hiring machinery would not only be a collosal waste of resources, but would lead to sub-optimal talent acquisition in the organisation, negatively impacting the organisation's productivity and performance. Recruitment drives differ from each other depending on the recruiting organisation, its size, and mode of selection (online / face-to-face, written / interview / both, single-step / multi-step etc.). Hence, while talking about improving the efficiency and effectiveness of the talent hiring process, it is difficult to talk about talent hiring in general. 

In this paper, we address a specific form of hiring: large recruitment drives involving testing/interviewing hundreds or thousands of candidates. This is a common scenario in case of large IT firms. Often, requirements stream in corresponding to both technology horizontals (Java, C, web programming, database administrator etc.) and business verticals (insurance, banking, aerospace, healthcare etc.). Shortlisting happens from tens of thousands of applications, and the shortlisted candidates are examined through possibly multiple rounds involving written tests and/or interviews.

Such recruitment drives cost a lot to organisations because of the infrastructural resources and because a large number of subject matter experts (SMEs) have to take time off from their regular work to man the selection panels. Given this, such recruitment drives should be done efficiently. Efficiency of such recruitment drives can be measured in many ways, e.g. as the ratio of resource expenditure to the number of reviews/interviews happening, or the number of successful recruitments.

Effectiveness of selection processes is a less tangible, but more important, parameter. Ideally, effectiveness can only be measured by tracking the contributions of the selected candidates through longitudinal studies. Such an ideal approach is hard to implement in practice. However, we believe that one of the elements that would significantly influence the effectiveness of any selection process is the extent to which the competence of a candidate is matched with those of the reviewers/interviewers who examine him/her. The reason behind this is that if a candidate with certain purported skills is reviewed by an SME in relevant areas, the review is likely to be deeper and more effective. Contrary to this, if the skills of a candidate and the reviewer are ill-matched, there will be mistakes made both ways: bad candidates may get through while the good ones do not get a fair chance.

In this paper, we investigate two important sub-problems of a recruitment drive: \emph{review/interview panel creation} and \emph{interview scheduling}, clubbed together as \emph{application processing problem}. In panel creation \footnote{review/interview panel creation} the goal would be to maximise the efficacy of the recruitment process by constituting panels such that the skill match between candidates and examiners is maximised. Interview scheduling aims to ensure that interviews are conducted as much in parallel as possible without causing conflicts. This leads to a prompt completion of the process thus saving time for everyone involved.

As regards the current state-of-practice (which is predominantly manual) the cost considerations in application processing emerge from the following sources:
\begin{enumerate}
\item Panel creation and scheduling are very costly and tedious processes.
\item The results of manually doing application processing lead to suboptimal utilisation of resources both in panel creation and in interview scheduling.
\end{enumerate}

Our approach consists of coming up with multiple algorithms for solving panel creation and interview scheduling problems automatically and experimentally comparing their performance based on efficiency and efficacy metrics which we introduce in this paper. We wish to situate our work in the context of recruitment drives conducted by large corporate organisations. Due to the paucity of data from such sources, our experiments have been carried out in the setting of application processing in an academic institution. Although this setup is much smaller than that of corporate recruitment drives and may differ in some details, it has all the essential properties of application processing as applied to corporate recruitment drives. Our experiments show strong evidences that automating the application processing process not only relieves its executors of the tedium, it also produces significantly more effective panels and efficient schedules than possible through manual process. The benefits of automation are expected to get more prominent as the scale increases. Our experiments, however, do not point towards one clear winner among the many algorithmic options.

Our work makes the following specific contributions:
\begin{enumerate}
\item mathematical formulation of panel creation and interview scheduling problem
\item algorithms for automatically solving the above two problems
\item metrics for estimation of efficiency and efficacy of application processing
\item experimental evaluation of various alternatives.
\end{enumerate}

Both the individual problems map to well-known problems in computing. To these, a lot of algorithmic solutions have been proposed in research literature. However, this paper is the first work as per our knowledge, which situates these problems in the application processing context. In that sense, this research is not a contribution to the field of algorithms. Instead, the value of our work is in identifying the algorithmic aspects of a specific real-world problem (application processing) and providing rigorous experimental evaluation of available solution alternatives. Most of these algorithms involve contextualisation of existing algorithms to the application processing problem. The findings of this work may be directly useful to solution providers in this space.

The rest of this paper is organised as follows: we present the application processing problem in detail in section~\ref{s:problem}. In section~\ref{s:approach}, we present in detail our scheme for automating the two aspects of application processing, namely panel creation and interview scheduling. In section~\ref{s:exp}, we present an overview of our experimental setup, the criteria by which the performance of the proposed algorithms is measured and the experimental results. We discuss related work in section~\ref{s:rw} and conclude the paper in section~\ref{s:conc}.

\section{Application Processing Problem} \label{s:problem}
\subsection{Illustrative Setup}
We introduce the application in the context of an academic university/institution. Consider an academic institute with around 50-60 faculty members working in 6-7 research domains (e.g. computer science, software engineering, networking and communication etc.). The institution has a student population of about 1000, out of which about 100 are research students. Anywhere between 500-1000 online applications come in for research programmes (Ph.D. and M.S. by research) every admission cycle, out of which about 10-15\% are made offers. A subset of all the applications are shortlisted through a review of applications. Final selection happens through an interview between the shortlisted candidate and a panel of faculty members. Please note that this example is of an institution of a small size compared to most institutions of higher learning in India.

Each application is reviewed, and each short-listed candidate interviewed, by a panel of 3-4 faculty members. These faculty members are chosen based on the overlap between the candidate's research interests and the faculty's. These panels (both for application review and interview) are created manually by an official who tries to make an informed guess to this effect. Note that finding matching research interests manually is far from straight forward. In these days of multi-disciplinary research, areas hardly ever settle into neat hierarchies. For example, a person interested in machine-learning (domain: data-science) may want to apply them to governance (domain: IT and society). This makes manual sorting of applications complex and error-prone. The ramifications of such errors are serious: research candidates with atypical research interests may end up getting reviewed by faculty members who are not able to judge their interests with a holistic perspective. In turn, this affects the quality of review (both at shortlisting and interview level) leading to errors in both directions: undeserving candidates may get in while deserving ones are missed.

Assume that panels are created with exclusive attention paid to overlap in research interests. This leads to another difficult issue. Panelists may end up in an arbitrary combination of panels. Pairs of panels with at least one panelist in common are said to conflict with each other. Interview schedule creators are under pressure to schedule interviews as much in parallel as possible so that the timespan of the entire interview process can be as small as possible. However, no two panels running in parallel must conflict. With hundreds of interviews to be conducted, detecting conflicts between panels is a very difficult task.

It turns out that both the above problems -- panel creation and interview scheduling -- map to well-known computing problems. In the remainder of this section, we present a mathematical formulation of both the problems. In section ~\ref{s:approach}, we present our approach to automatically solve these problems, both using algorithms designed by us, and using variants of existing algorithms.

\subsection{Panel Creation} \label{s:probpc}
Consider a set $P$ of panelists, a set $C$ of candidates and a set $T$ of topics of interest.
We define \emph{an assignment} $G$ as a bipartite graph. The left column of $G$ -- given by $panelists(G)$ -- consists of nodes, each of which corresponds to one of the panelists, and the right column -- given by $candidates(G)$ -- consists of nodes, each of which corresponds to one of the candidates. An edge $e_{ij}$ will run between a panelist $p_i$ and candidate $c_j$ if there is at least one topic of interest common between $p_i$ and $c_j$. $e_{ij}$ will be annotated with the set of topics which are common between $p_i$ and $c_j$.

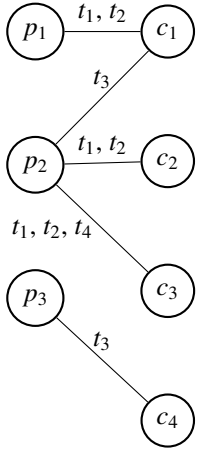
\begin{figure}[t]
\begin{center}
\begin{tikzpicture}
\node[circle, draw=black, thick](p1){$p_1$};
\node[circle, draw=black, thick, below=of p1](p2){$p_2$};
\node[circle, draw=black, thick, below=of p2](p3){$p_3$};

\node[circle, draw=black, thick, right=of p1](c1){$c_1$};
\node[circle, draw=black, thick, below=of c1](c2){$c_2$};
\node[circle, draw=black, thick, below=of c2](c3){$c_3$};
\node[circle, draw=black, thick, below=of c3](c4){$c_4$};

\draw[-] (p1) to node[above]{$t_1$, $t_2$}(c1);
\draw[-] (p2) to node[above]{$t_3$}(c1);
\draw[-] (p2) to node[above]{$t_1$, $t_2$}(c2);
\draw[-] (p2) to node[left]{$t_1$, $t_2$, $t_4$}(c3);
\draw[-] (p3) to node[above]{$t_3$}(c4);
\end{tikzpicture}
\end{center}
\caption{Example: Panelists $p_1$, $p_2$, $p_3$, Candidates $c_1$, $c_2$, $c_3$, $c_4$, Topics  $t_1$, $t_2$, $t_3$, $t_4$}
\label{f:pc-example}
\end{figure}

\begin{example}
Consider the example shown in Figure~\ref*{f:pc-example}. We have three panelists $p_1$, $p_2$, $p_3$, four candidates $c_1$, $c_2$, $c_3$, $c_4$. Each edge between a panelist and candidate is annotated with one or more topics. Here, the complete set of topics is \{$t_1$, $t_2$, $t_3$, $t_4$\}. Edges are $e_1: (p_1, c_1, \{t_1, t_2\})$, $e_2: (p_2, c_1, \{t_3\})$, $e_3: (p_2, c_2, \{t_1, t_2\})$, $e_4: (p_2, c_3, \{t_1, t_2, t_4\})$, $e_5: (p_3, c_4, \{t_3\})$.
$\blacksquare$
\end{example}

A panel $\mathcal{P}$ is a pair $(candidate, panel)$ where $candidate$ is a candidate, and $panel : 2^P$ is a set of panelists. A paneling $\mathbb{P} : C \mapsto 2^{P}$ is a map from one candidate in $C$ to a subset from $P$. If, for a panel $\mathcal{P}$, $\mathcal{P}.candidate \in dom(\mathbb{P})$ and  $\mathcal{P}.panel = \mathbb{P}(\mathcal{P}.candidate)$, then we say, $\mathcal{P} \in \mathbb{P}$. 

We define the \emph{value} $V$ of a panel $\mathcal{P}$ as a function -- currently not specified -- that maps it to a real value representing the quality of a panel. The \emph{cummulative value} $\mathcal{V}$ of $\mathbb{P}$ is the sum of the values of its individual panels.
\begin{equation}
\mathcal{V}(\mathbb{P}) = \sum\limits_{\mathcal{P} \in \mathbb{P}} V(\mathcal{P})
\end{equation}

The goal of panel creation is to find a $\mathbb{P}$ such that $\mathcal{V}(\mathbb{P})$ is the maximised. 

\subsection{Interview Scheduling} \label{s:probis}

\subsubsection{Conflicting Panels}
Consider a set of interview panels $\mathbb{P}$ that have been computed by solving the problem presented in Section ~\ref{s:probpc} using one of the algorithms presented in Section~\ref{s:apppc}. We say that two panels $\mathcal{P}_i, \mathcal{P}_j \in \mathbb{P}$ conflict with each other if there is at least one panelist common to them.
\begin{equation}
conflict(\mathcal{P}_i, \mathcal{P}_j) = \mathcal{P}_i.panel \cap \mathcal{P}_j.panel \neq \phi
\end{equation}

\begin{example}
Let $P = \{p_{1}, p_{2}, p_{3}, p_{4}, p_{5}, p_{6}, p_{7}, p_{8}, p_{9} \}$ be the set of panelists. Let us assume we have generated six panels using one of the panel creation algorithms, and each panel consists of one candidate and a set of panelists. Here, the panels generated are: $\mathcal{P}_1 = (c_1, \{p_1, p_2, p_3\})$, $\mathcal{P}_2 = (c_2, \{p_3, p_4, p_5\})$, $\mathcal{P}_3 = (c_3, \{p_5, p_6, p_7\})$, $\mathcal{P}_4 = (c_4, \{p_1, p_2, p_6\})$, $\mathcal{P}_5 = (c_5, \{p_7, p_8\})$ and $\mathcal{P}_6 = (c_6, \{p_9, p_{10}\})$. Here, we can see that $\mathcal{P}_1$ and $\mathcal{P}_2$ conflict (i.e. $conflict(\mathcal{P}_1, \mathcal{P}_2) = true$) because their panels have a common panelist $p_3$. Similarly, there are other conflicts pairs of panel, e.g. $\mathcal{P}_1$ and $\mathcal{P}_4$ etc.
$\blacksquare$
\end{example}

\subsubsection{Valid Schedule}
An interview is scheduled for each candidate. Assuming that panel creation step has been completed, each candidate has a interview panel associated with it. A schedule $S$ maps each interview to a time-slot or interval, i.e. $S : \mathbb{P} \rightarrow interval$. An $interval$ $i$ is a pair $(s, e)$ where $i.s$ is starting time and $i.e$ is the ending time of $i$. Two intervals (or time slots) $i_1$ and $i_2$ are said to overlap if either starts before the other ends, i.e. $overlap(i_1, i_2) = (i_2.e > i_1.s > i_2.s) \lor  (i_1.e > i_2.s > i_1.s)$. A schedule is valid or feasible if no two conflicting panels are mapped to overlapping slots, i.e.

\begin{multline}
valid(S) = \\ \text{\hspace{1cm}}\forall \mathcal{P}_i, \mathcal{P}_j \in \mathbb{P},\ conflict(\mathcal{P}_i, \mathcal{P}_j) \implies \\ \text{\hspace{1cm}} \neg overlap(S(\mathcal{P}_i), S(\mathcal{P}_j))
\end{multline}

\subsubsection{Graph Colouring}
\emph{Interference graph.} We use the data of every panel $\mathcal{P} \in \mathbb{P}$  to create an interference graph $I$, where each node $n_{i}$ represents a panel $\mathcal{P}_i$, and an edge $e_{ij}$ exists between two such nodes $n_{i}$ and $n_{j}$ if $conflict(\mathcal{P}_i, \mathcal{P}_j) = true$. The weight $w_{ij}$ of the edges signify the number of common panelists.

Since an edge between two nodes indicates a panelist who is common to both the panels, these two panels cannot be scheduled simultaneously in a valid schedule. This means that if we colour a node $n_{i}$ with a specific colour $C_{i}$, which denotes a particular time interval in which the interview is to be scheduled for the panel denoted by $n_{i}$, we need to ensure that no two adjacent nodes have the same colour.  This leads us to classifying the interview scheduling process as \emph{graph colouring problem}, and we make use of one of the three algorithms mentioned in the subsequent sections to colour the graph accordingly.

\begin{example}
Consider the interference graph shown in Figure~\ref{f:exisch}. The edge $e_{12}$ with weight $w_{12} = 1$ denotes that there is one common panelist between the panels denoted by $c_{1}$ and $c_2$; edge $e_{14}$ with weight $w_{14} = 2$ denotes that there are two common panelists between the panels denoted by $c_{1}$ and $c_4$, and so on for edges $e_{23}, e_{34}$, and $e_{35}$.

The goal of the interview scheduling problem is to colour the graph in $k$ colours, such that $k$ is minimized. The minimum value of $k$ will be determined by one of the graph colouring algorithms. 
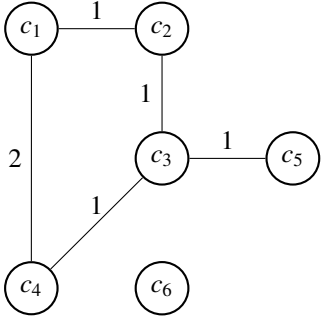
\begin{figure}[t]
\begin{center}
\begin{tikzpicture}
\node[circle, draw=black, thick](c1){$c_1$};
\node[circle, draw=black, thick, right=of c1](c2){$c_2$};
\node[circle, draw=black, thick, below=of c2](c3){$c_3$};
\node[circle, draw=black, thick, right=of c3](c5){$c_5$};
\node[circle, draw=black, thick, below=of c3](c6){$c_6$};
\node[circle, draw=black, thick, left=of c6](c4){$c_4$};

\draw[-] (c1) to node[above]{$1$}(c2);
\draw[-] (c2) to node[left]{$1$}(c3);
\draw[-] (c3) to node[above]{$1$}(c5);
\draw[-] (c1) to node[left]{$2$}(c4);
\draw[-] (c3) to node[above]{$1$}(c4);
\end{tikzpicture}
\end{center}
\caption{Example: Candidates $c_1$ to $c_6$}
\label{f:exisch}
\end{figure}
$\blacksquare$
\end{example}

\section{Our Approach} \label{s:approach}
\subsection{Panel Creation} \label{s:apppc}

\subsubsection{Edge sorting approach}

\begin{algorithm}
\begin{algorithmic}
\State $S_c(p)$ is defined in equation (\ref{e:score}) 
\State $Load(p)$: keeps track of load (number of candidates allotted) for each panelist\\
\Function{generatePanels}{$G$}
	\State $P, C \gets panelists(G), candidates(G)$
	\State $Load(p) \gets$ initialized to 0 for each panelist
	\ForAll{$c \in C$}
		\State $E \gets$ set of edges connected to candidate $c$
		\State $n \gets \mathbb{S}$
		\Comment{$\mathbb{S}$: Panel size}
		\State $m \gets \mathbb{L}$ 
		\Comment{$\mathbb{L}$: Maximum load of a panelist}
		\State $\mathbb{P}[c] \gets$ set of $n$ members of $E_c$ with maximum $S_c(p)$ and $Load(p) < m$
		\State $Load(p) \gets$ increment by 1 for each $p$ in $\mathbb{P}[c]$
	\EndFor
	\State \textbf{return} $\mathbb{P}$
\EndFunction
\end{algorithmic}
\caption{Panel creation -- Edge sorting approach}
\label{a:edge}
\end{algorithm}

The edge sorting approach is presented in Algorithm~\ref{a:edge}.
Panelists occupy the left column of $G$, given by $panelists(G)$, abbreviated as $P$; the candidates occupy the right column of $G$, given by $candidates(G)$ abbreviated as $C$. The idea is to try and ensure that a panelist with larger overlap with the interests of a candidate should preferably be one of the panelists for that candidate. We elaborate this idea as follows: 
\begin{enumerate}
\item More exclusive matches between a panelist and candidate should be given more preference as compared to generic matches. This means that of two panelists $p_1$ and $p_2$ who are connected to a candidate $c$ through the same set of topics, if $p_1$ has fewer other matches with other candidates as compared to $p_2$ (probably due to $p_1$'s area of expertise being specialised/rarer than that of $p_2$), then the match between $p_1$ and $c$ should be preferred over that between $p_2$ and $c$. This policy applies the other way too. That is, of two candidates $c_1$ and $c_2$ who are connected to a panelist $p$ through the same set of topics, if $c_1$ has fewer other matches with other panelists as compared to $c_2$, then the match between $c_1$ and $p$ should be preferred over that between $c_2$ and $p$.
\item More exclusive topics contribute more to the edge weight as compared to more generic topics. Suppose a candidate $c$ is connected to one panelist $p_1$ through a topic $t_1$ and to another panelist $p_2$ through a topic $t_2$. Assume that the scores of both the panelists are same. Which panelist should get higher preference? This is determined on the basis of how exclusive or generic $t_1$ and $t_2$ are. Suppose that the pair $(p_1, c)$ is the only edge which is annotated by $t_1$ (probably because $t_1$ is a very specialised topic in the given organisation, e.g. public policy and IT), while $t_2$ (may be a popular topic like machine learning or cybersecurity) sits on many other edges in the graph apart from $(p_2, c)$. Then $p_1$ would get more preference than $p_2$ as a panelist for $c$. 
\end{enumerate}

The above policies are mathematically summarised in the following equation:
\begin{equation}
S_{c}(p) = \left\{
        \begin{array}{ll}
            \frac{\sum\limits_{t \in T(p, c)}\frac{1}{S_t}}{S_p} & T(p,c) \neq \phi \\
            -\frac{1}{1+S_p} & T(p,c) = \phi
        \end{array}
    \right.
\label{e:score}
\end{equation}

In the above, $S_{c}(p)$ denotes the overall match score for panelist $p$ w.r.t. a candidate $c$. $T(p, c)$ gives the set of topics common to $p$ and $c$, i.e. $T(p, c) = T(p) \cap T(c)$ where $T(p)$ and $T(c)$ are the sets of topics of interest of panelist $p$ and candidate $c$ respectively. Panelist score $S_p$, candidate score $S_c$ and topic score $S_t$ are the number of edges panelist $p$, candidate $c$ and topic $t$ appear on respectively in the bipartite graph $G$. The intuition behind the above equation are as follows:
\begin{itemize}
\item The term in the numerator $\sum\limits_{t \in T(p, c)}\frac{1}{S_t}$ is the sum of reciprocals of scores of all topics involved between $p$ and $c$. This summarises the intuition that with an increase in the genericness of a topic $t$ (quantified by its score $S_t$) its contribution to the strength of the connection between a panelist and candidate should reduce. However, with more topics associating two individuals, the strength should go up. Hence, the summation.
\item The score of a panelist w.r.t. a candidate would be inversely proportional to the genericness of a panelist (quantified by his/her score $S_p$). Hence, it appears as the denominator of the expression.
\item A panelist may have to be included in the panel of a candidate even though they may share no common interest in case the number of panelists who share common interests with the candidate is not enough to form the panel. In that case, we assign the score $-\frac{1}{1+S_p}$, which will help us to assign the more generic panelist and at the same time panelist which have some common topics of interest with the candidate still get priority over the panelist who does not.
\item If a panelist shares no common interest with the candidate, there is also a possibility where that panelist does not share common interest with any candidate. $S_p$ for that panelist would be $0$. An offset of $1$ is included in denominator to handle this case.

\end{itemize}

The panel for a candidate $c$, $\mathbb{P}(c)$ is computed by selecting the set of edges whose scores are the highest ensuring that $|\mathbb{P}| = \mathbb{S}$ and $Load(p) < \mathbb{L}$ where $\mathbb{S}$ is the size of the panel and $\mathbb{L}$ is the maximum allowed load for a panelist. 

\begin{example}

\begin{figure}[t]
\begin{center}
\begin{tikzpicture}
\node[circle, draw=black, thick](p1){$p_1$};
\node[circle, draw=black, thick, below=of p1](p2){$p_2$};

\node[circle, draw=black, thick, right=of p1](c1){$c_1$};
\node[circle, draw=black, thick, below=of c1](c2){$c_2$};

\draw[-] (p1) to node[above]{$t_1$, $t_2$}(c1);
\draw[-] (p1) to node[above,rotate =-45]{$t_1$, $t_2$,$t_3$,$t_4$}(c2);
\draw[-] (p2) to node[above,pos = 0.1,rotate=45]{$t_1$}(c1);
\draw[-] (p2) to node[above]{$t_1$}(c2);
\end{tikzpicture}
\end{center}
\caption{Example: Edge Sorting Algorithm}
\label{f:edge-example}
\end{figure}

Panelist weights: $W(p_1) = 2$, $W(p_2) = 2$.

Topic weights: $W(t_1) = 4$, $W(t_2) = 2$, $W(t_3) = 1$, $W(t_4) = 1$.

Taking $\mathbb{S} = 1$ and $\mathbb{L}=1$

Edge weights:
\begin{itemize}
\item $W(e_1) = S_{c_1}(p_1) = \frac{\frac{1}{W(t_1)} + \frac{1}{W(t_2)}}{W(p_1)} = \frac{\frac{1}{4} + \frac{1}{2}}{2} = 0.375$ 
\item $W(e_2) = S_{c_1}(p_2) = \frac{\frac{1}{W(t_1)}}{W(p_2)} = \frac{\frac{1}{4}}{2} = 0.125$
\item $W(e_3) = S_{c_2}(p_1) = \frac{\frac{1}{W(t_1)} + \frac{1}{W(t_2)} + \frac{1}{W(t_3)} + \frac{1}{W(t_4)}}{W(p_1)} = \frac{\frac{1}{4} + \frac{1}{2} + \frac{1}{1} + \frac{1}{1}}{2} = 1.375$
\item $W(e_4) = S_{c_2}(p_2) = \frac{\frac{1}{W(t_1)}}{W(p_2)} = \frac{\frac{1}{4}}{2} = 0.125$
\end{itemize}

We compute panels for each candidate using Edge Sorting Algorithm by following steps:
\begin{enumerate}
    \item For candidate $c_1$, we have two options for panelists - $\{ p_1,p_2 \}$ with $Load(p) < 1$ (as load is $0$ for every panelist initially). 
    \item Out of these, we will assign $1$ panelist to $c_1$ with maximum score, i.e. $p1$. Therefore, $\mathbb{P}[c_1] = \{p_1\}$.
    \item Now we increment the load of $p1$ by 1. Hence, $Load(p_1)=1$.
    \item For candidate $c_2$, we have only one option for panelist - $\{ p_2 \}$ with $Load(p) < 1$.
    \item So, we assign this panelist to $c_1$. Therefore $\mathbb{P}[c_2] = \{p_2\}$.
\end{enumerate}
Total score of assignment becomes $0.375+0.125 = 0.5$.
$\blacksquare$
\end{example}

\subsubsection{Min-Cost-Max-Flow Approach}
\begin{algorithm}
\begin{algorithmic}
\Function {MinCostMaxFlow}{$G'$, $s$,$t$}
    \ForAll{Edges $E(u,v) \in G'$}
        \State Add edge, $e$ directed from $v$ to $u$  
        \State $Capacity[v][u] \gets 0$
        \State $Weight[u][v] \gets -Weight[u][v]$
        \State $Weight[v][u] \gets - Weight[u][v]$
        \State $Flow[u][v] \gets Flow[v][u] \gets 0$
    \EndFor
    
    \While{$\exists$ path from $s$ to $t$ in $G'$}
        \State $p \gets$ Shortest Path from $s$ to $t$
        \State $f \gets min(Capacity(\forall$ Edges $\in$ path $p$) ) 
        \ForAll{Edges $E(u,v) \in$ path $p$}
            \State $Capacity[u][v] \gets Capacity[u][v] - f$
            \State $Capacity[v][u] \gets Capacity[v][u] + f$
            \State $Flow[u][v] \gets Flow[u][v] + f$
        \EndFor
    \EndWhile
    \State \textbf{return} $Flow$
\EndFunction
\Statex
\Function{generatePanels}{$G$}
    \State $G' \gets \Call{augmentGraph}{G}$ as follows:
    \State $FlowMap \gets$ MinCostMaxFlow($G'$,$s$,$t$)
    \State $P, C \gets panelists(G), candidates(G)$
    \ForAll{$c \in C$}
        \ForAll{$p \in P$}
            \If{ FlowMap[p][c] - FlowMap[c][p]  = 1}
                \State $\mathbb{P}[c] \gets$ add panelist $p$ 
            \EndIf
        \EndFor
    \EndFor
    \State \textbf{return} $\mathbb{P}$
\EndFunction

\Statex
\Function{augmentGraph}{$G$} %\Comment{Example: Figure~\ref{f:gaug}}
    \State Create a graph $G'$ isomorphic to $G$ except the following:
    \begin{enumerate}
    \item Node $s$ (Source Node) with edges directed from $s$ to Panelist Nodes with
            edge capacity \textit{$\mathbb{L}$} and weights $0$.
    \item Node $t$ (Sink Node) with edges directed from Candidates Nodes to $t$ 
        with  edge capacity \textit{$\mathbb{S}$} and weights $0$.
    \end{enumerate}
    
	\State \textbf{return} $G'$
\EndFunction
\end{algorithmic}
\caption{Panel creation -- Min Cost Max Flow Approach }
\label{a:flow}
\end{algorithm}

\begin{figure}
\begin{tikzpicture}
\node[circle, draw=Black](p11) at (0, 0){$P_1$};
\node[below=0.1cm of p1](d11) {$\vdots$};
\node[circle, draw=Black, below=0.1cm of d11](p21){$P_2$};
\node[circle, draw=Black, right=of p11](c11){$C_1$};
\node[below=0.1cm of c11](d21) {$\vdots$};
\node[circle, draw=Black, below=0.1cm of d21](c21){$C_2$};

\node[right=0.5 of d21](a){$\Rightarrow$};

\node[circle, draw=Black, right=2 of c11](p12) {$P_1$};
\node[below=0.1cm of p12](d12) {$\vdots$};
\node[circle, draw=Black, below=0.1cm of d12](p22){$P_2$};
\node[circle, draw=Black, right=of p12](c12){$C_1$};
\node[below=0.1cm of c12](d22) {$\vdots$};
\node[circle, draw=Black, below=0.1cm of d22](c22){$C_2$};

\node[circle, draw=Black, right=0.5 of a](s) {$s$};
\node[circle, draw=Black, right=3 of s](t) {$t$};

\draw[->](s) -- (p12);
\draw[->](s) -- (p22);
\draw[->](c12) -- (t);
\draw[->](c22) -- (t);
\end{tikzpicture}
\caption{Graph augmentation for Min Cost Max Flow algorithm}
\label{f:gaug}
\end{figure}
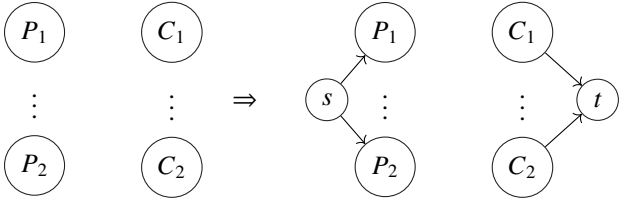

The Minimum Cost Maximum Flow approach is presented in Algorithm~\ref{a:flow}.

Input problem instance is modeled as a bipartite graph $G$ in exactly same way as it is done in edge sorting approach. The weights of the edges are also defined as before.\\

In this approach, the concept of \textit{Network Flows} \cite{network_flows} is used to solve the given problem. We mapped the given problem to a well known problem in this field, \textit{Minimum Cost Maximum Flow Problem} \cite{mincostmaxflow_algo}.
Various algorithms have been derived to solve this problem optimally. We will use \textit{Min-Cost-Max-Flow Algorithm} \cite{mincostmaxflow_algo} as part of our solution to given problem.\\

In bipartite graph $G$, for every candidate in $candidates(G)$, a panelist from $panelists(G)$ has to be assigned such that:
\begin{itemize}
\item every candidate is assigned a panel whose size is $\mathbb{S}$;
\item each panelist is assigned at most $\mathbb{L}$ number of candidates
\item Overall cost of assignment is maximized.
\end{itemize}

For mapping given problem to \textit{Minimum Cost Maximum Flow Problem}, given complete bipartite graph $G$ of original problem is augmented to a graph $G'$ as given in \textit{AugmentGraph(G)} (Algorithm \ref{a:flow}. \textit{Min-Cost-Max-Flow Algorithm} is then run on this augmented graph. The desired solution is extracted from the optimal flow we get. We elaborate this idea as follows:

\begin{enumerate}
    \item Since each panelist can be assigned to at most $\mathbb{L}$ number of candidates, we can think of panelists having $\mathbb{L}$ tickets each that they can give to candidates. That is why edge capacity of $\mathbb{L}$ has been kept from source node to panelist node. 
    
    \item Each panelist can give at most 1 ticket to a candidate. Giving no ticket or $0$ ticket would mean that the panelist is not assigned to the given candidate and giving $1$ ticket would mean that the candidate is assigned to the panelist. That is why edge capacity of $1$ has been kept from panelist node to candidate node.
    
    \item Now, each candidate has to be assigned $\mathbb{S}$ panelists. This can be inferred as each candidate receiving $\mathbb{S}$ tickets. Since each panelist can give at most one ticket to a candidate, receiving $\mathbb{S}$ tickets by a candidate would mean getting $\mathbb{S}$ different panelists. That is why edge capacity of $\mathbb{S}$ has been kept from candidate node to sink node.
    
    \item Now, suppose that there are \textit{Number Of Panelists} $\times$ $\mathbb{L}$ tickets present at source node. In the case of maximum flow, each panelist would be getting at most $\mathbb{L}$ tickets from source node, given the capacity of source to panelist edges. 
    
    \item These tickets would be further allotted to the candidates by the respective panelists. Assuming there are sufficient number of panelists and value of $\mathbb{L}$, in case of maximum flow, all the edges from candidates to sink node would have flow equal to their capacity. Hence all candidates will always have exactly $\mathbb{S}$ number of panelists assigned.
    
    \item The cost assigned to edges originating from source node and edges ending at sink node is $0$, because they have no role to play in final cost of assignment we want.
    
    \item As we also have to find the optimal cost of assignment, we have to get maximum flow with the optimal cost. In our case, the cost has to be maximized. However there is no direct algorithm that can find maximum flow in a graph with maximum cost. Hence, we change the signs of our original weights ($S_c$($p$) to $-S_c$($p$)) so that when algorithm minimizes cost with negative weights, the original cost with positive weights get maximized.
    
    \item Finally, when we get optimal flow after performing the algorithm, we can assign the panelist to those candidates whose net flow is $1$.

\end{enumerate}

\textbf{Note:} Having the edge capacity 0 would not allow the flow from that edge, hence it is equivalent to saying that the path does not exist from that edge. Since the capacity of edge is reduced by flow value ($f$) each time flow is augmented along the shortest path, there will be some point when the path would not exist from source to sink node.

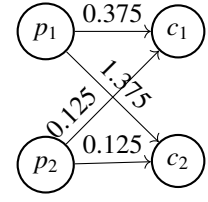
\begin{figure}[t]
\begin{center}
\begin{tikzpicture}
\node[circle, draw=black, thick](p1){$p_1$};
\node[circle, draw=black, thick, below=of p1](p2){$p_2$};

\node[circle, draw=black, thick, right=of p1](c1){$c_1$};
\node[circle, draw=black, thick, below=of c1](c2){$c_2$};

\draw[->] (p1) to node[above]{$0.375$}(c1);
\draw[->] (p1) to node[above,rotate =-45]{$1.375$}(c2);
\draw[->] (p2) to node[above,pos = 0.2,rotate=45]{$0.125$}(c1);
\draw[->] (p2) to node[above]{$0.125$}(c2);
\end{tikzpicture}
\end{center}
\caption{Example: Min-Cost-Max-Flow Algorithm}
\label{f:maxflow-example}
\end{figure}

\begin{example}

\begin{figure*}
\includegraphics[width=0.9\textwidth]{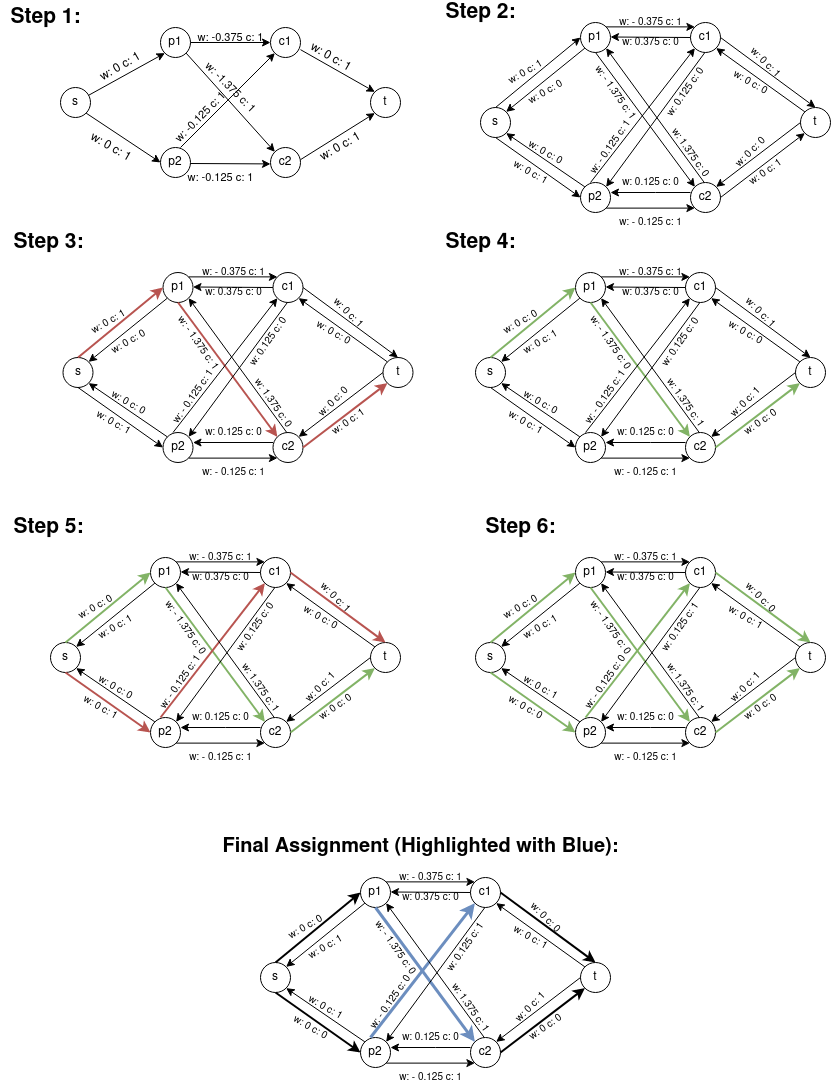}
\caption{Min Cost Max Flow Example}
\label{f:maxflow-example}
\end{figure*}
Let us consider the same example from the last section. The annotations over edges represents the weights assigned as derived before. We compute panels for each candidate using Min-Cost-Max-Flow Algorithm by following steps (shown pictorially in Figure~\ref{f:maxflow-example}):
\begin{enumerate}
    \item For representation purposes, we will be annotating the edges with $w$ representing the weight of the edge and $c$ representing the capacity of edge. If the edge is having flow of 1 unit, it is highlighted with green colour.
    
    \item Firstly, the given graph is augmented. (Step 1 in figure)
    
    \item Min Cost Max Flow algorithm is applied on this graph. Min Cost Flow algorithm first does some preprocessing on the graph on which it is applied. (Step 2 in figure)
    
    \item Then we find the shortest path from source node to sink node in above graph. Highlighted part with red colour in resulting graph represents the only shortest path possible. (Step 3 in figure)
    
    \item We get the following residual graph after augmenting the flow along this path. (Step 4 in figure)
    
    \item Again, we find the shortest path from source node to sink node in updated graph. Highlighted part with red colour in resulting graph represents the only shortest path possible. (Step 5 in figure)
    
    \item We get the following residual graph after augmenting the flow along this path.(Step 6 in figure)
    
    \item We can see that there is no path possible from source node to sink node in the updated graph. Hence we exit the while loop.
    
    \item Now, when we find panelist to candidate assignments in original graph for getting the panels created with help of above network flow, we get final assignment shown highlighted with blue colour. (Step 7 in figure) 
    
    \item Hence, $\mathbb{P}[c_1]=\{ p_2 \}$ and $\mathbb{P}[c_2]=\{ p_1 \}$.
\end{enumerate}

Total score of assignment becomes $1.375 + 0.125 = 1.5$.
$\blacksquare$
\end{example}

\subsubsection{Proof of Optimality}

It has been proven that the given \emph{Min-Cost-Max-Flow Algorithm} gives an optimal solution for \emph{Minimum Cost Maximum Flow Problem} in Network Flows. As we successfully mapped this problem to given problem, we can say that we get an optimal solution for the given problem.

\subsubsection{Comparison with Edge Sorting Approach}

In the \emph{Edge Sorting} algorithm, we can intuitively observe that a greedy approach is followed. For every candidate, we are choosing some best panelists based on the score between the candidate and panelist. This is somehow locally optimizing the problem and does not guarantee an overall optimized solution. To support this statement, we can also see the results of running both algorithms on same example. When we ran \textit{Edge Sorting Algorithm}, the overall cost of assignment we got was $0.125$, whereas, for same example, \textit{Min-Cost-Max-Flow} approach gave total cost of assignment = $1.5$. However, it is trivial to notice that if we would have started \textit{Edge sorting Algorithm} from candidate 2 ($c_{2}$), we would have got total cost of assignment = $1.5$ which is same as latter approach.

Although, following the greedy approach seems easier to understand and implement, we felt a need to come up with a deterministic optimal approach for this problem. This approach is a bit less intuitive to understand and implement than the previous approach but it guarantees overall optimal solution always.

\subsection{Interview Scheduling} \label{s:appis}

In this subsection, we discuss three approaches to solve the interview scheduling problem. As mentioned earlier (see Section~\ref{s:probis}), we model this problem as the graph colouring problem. All three algorithms presented here really solve graph colouring problem to create the interview schedule.

\subsubsection{Approach 1 -- Chaitin's algorithm}
\begin{algorithm}
\begin{algorithmic}
\Function{chaitin}{$G$, $k$}
\State Stack $S \gets \phi$
\While{$G$ is not empty}
	\If{$\exists$ node $n$ in $G$ such that $degree(n) \le k$}
		\State Remove $n$ from $G$ along with all edges connected to $n$.
		\State \Call{push}{$n$, $S$}
	\Else
		\State Fail or restart with a larger value of $k$.
	\EndIf
\EndWhile
\While{$S$ is not empty}
	\State $n \gets$ \Call{pop}{$S$}
	\State Restore $n$ to $G$ along with the connected edges.
	\State $C[n] \gets$ $c$, the least colour which not the colour of any of $n's$ current neighbours. 
\EndWhile
\State \textbf{return} $C$
\EndFunction
\end{algorithmic}
\caption{Interview scheduling -- Chaitin's algorithm}
\label{a:chaitin}
\end{algorithm}

The first approach, Chaitin's algorithm is shown in Algorithm~\ref{a:chaitin}. We will consider the graph shown in Figure~\ref{f:exisch}. In this approach, all the panels are represented as nodes in the graph $G$, and each panel will have one candidate and one or more panelists. 

Our graph $I$ can be represented as a dictionary as $I$: \{$c_1$ : [$c_2$, $c_4$], $c_2$ : [$c_1$, $c_3$], $c_3$ : [$c_2$,$c_4$,$c_5$], $c_4$ : [$c_1$, $c_3$], $c_5$ : [$c_3$], $c_6$ : []\}. Here, each candidate is mapped to other candidates who share a common panelist with the candidate. The various steps which the graph colouring algorithm goes through are listed below:
\begin{enumerate}
    \item The first step is to find a natural number $k$ such that we can push the nodes with a degree less than $k$ onto the stack. We make use of binary search to find the value of $k$. 
    \item Chaitin's algorithm is then run for that particular value of $k$ that was obtained in the first iteration (this value would be the mid point of the range). Now the nodes with a degree less than $k$ are pushed onto the stack $S$, and the node along with its edges will be removed from $G$. If it reaches a point where $G$ does not have any nodes with degree $\le$ $k$, we break the Chaitin's algorithm and run it again with a larger value of $k$ we get from the next iteration of our binary search.
    \item The above process is repeated until $G$ is empty and all the nodes are pushed onto the stack. Then we pop all the nodes that were stored in $S$, and restore them in $G$ along with its edges. While restoring the node onto the graph, we also give it a parameter called $colour$, which is taken from the $k$ of the Chaitin's algorithm, the range of $colour$ varying from 0 to $k-1$. The node that is popped is assigned the least value of $colour$, provided the value of its $colour$ is not shared by any of its adjacent nodes.
    \item At this step all the nodes are popped from $S$, and are placed back on the graph, and have also been assigned a colour. The minimum number of colours required to colour the graph would be the last valid value of $k$. 
    \item Regarding our scheduling problem, $k$ acts as the minimum number of slots required to conduct all the interviews. And since each node represents a panel, all the nodes which were coloured with the same value of $colour$ correspond to interviews that can be held simultaneously without any conflict with the other interviews corresponding to the same colour. For Figure~\ref{f:exisch}, we can compute its graph colouring by following the steps above, and the minimum number of colours required to colour it would be 2. And the graph colouring would be : $c_1 \mapsto 0$, $c_2 \mapsto 1$, $c_3 \mapsto 0$, $c_4 \mapsto 1$, $c_5 \mapsto 1$,$c_6 \mapsto 0$.
    \end{enumerate}

\subsubsection{Approach 2 -- Genetic algorithm}
\begin{algorithm}
\begin{algorithmic}
\State List $chromosomes \gets$ random colourized graphs
\Function{genetic}{$G$, $k$}
% \While{$generations \ne 0$}
\For{$i = 0$ to $noOfGenerations$}
    \State $p_1$, $p_2$ $\gets$ two parents from $chromosomes$ 
    \State $child$ $\gets$ \Call{crossover}{$p_1$, $p_2$}
    \State $child$ $\gets$ \Call{mutate}{$child$}
    \If{colour conflicts of $child$ \textless \ $p_1$ or $p_2$} 
    % Doubt : should we make this conflicts or fitness
        \State \Call{push}{$child$, $chromosomes$}
        \State \Call{pop}{$more\ conflicting\ parent$}
    \EndIf
\EndFor
\State \textbf{return} $child$, $child.conflicts$
\EndFunction
\Function{compare}{$G$, $k$}
\If{colour conflicts of $child$ is zero}
    \State \Call{genetic}{$G$, $k+1$}
\Else
    \State \Call{genetic}{$G$, $k-1$}
\EndIf
\EndFunction
\end{algorithmic}

\caption{Interview scheduling -- Genetic algorithm}
\label{a:genetic}
\end{algorithm}

The second algorithm, Genetic-Chaitin approach is shown in Algorithm~\ref{a:genetic}. Let us consider the graph shown in Figure~\ref{f:exisch} as an example again. In this approach, we combine Chaitin's and Genetic algorithm. Since Genetic algorithm needs a base of minimum colours on which it tries to improve, we use the minimum number of colours obtained from Chaitin's algorithm and set the Genetic algorithm to work on it to decrease the minimum number of colours required to colour the graph.

Let us consider each chromosome as a dictionary with each key denoting a \emph{candidate} whose value denotes a number corresponding to the colour that the \emph{candidate’s} node has been coloured with. This colour is generated at random initially for all the nodes.

The first step is to select two parents from the pool of chromosomes, and choose a crossover point in both which results in a child inheriting some colouring from the first parent as well as the second parent. The child is then mutated to achieve a better colouring, and then the two fittest among the parents and the child are chosen, the remaining chromosome is discarded from the population.

Taking the example of the graph shown in Figure~\ref{f:exisch}, we initially consider $chromosomes$ as a collection of randomly coloured graphs (50, as an example), where each graph is a copy of our original graph in Figure~\ref{f:exisch}. For each generation (which is an iteration of the entire process), we perform the following steps:
\begin{enumerate}
    \item We select any two parents from the $chromosomes$ using a linear combination of two methods:
    \begin{itemize}
        \item The first method selects the two parents randomly from the given population
        \item The second method selects the two fittest parents from the entire population (i.e the two parents having the least conflict between them)
    \end{itemize}
    \item After selecting the parents, we create a child by crossing-over the two parents using a random crosspoint. The first part of the child (until the crosspoint) comes from the first parent, and the second part (after the crosspoint) comes from the second parent. This process is called \emph{crossover}.
    \item After the child has been created, we \emph{mutate} the child (colouring the graph), albeit in a sub-optimal way. To mutate the child, we sort the nodes of the graph in descending order of their degrees, and then use either of the following two methods to colour the graph:
    \begin{itemize}
        \item The first method colours the adjacent nodes of all the nodes with a $valid\ colour$, where a $valid\ colour$ is any colour that can be used to colour a node while preserving the graph colouring property. 
        \item The second method colours the adjacent nodes of all the nodes with a random colour.
    \end{itemize}
    After trying both of the above functions on a collection of data sets, it is observed that the first function outperformed the second in most the cases.
    %Doubt to include this ihere or experiments?
    \item After the child has been mutated, we replace the more conflicting (less fit) parent from the chromosome list with the child  if and only if the conflicts in the child are less in number than the conflicts in both its parents.
    \item The above process (steps 1-3) is repeated for all the generations, after which we get a fittest child from the entire population. If the conflicts of the child is zero, it implies that the graph is coloured validly. And if there are more generations left after finding an optimal child, the algorithm tries to look for a much more fitter child (less number of colours required). Larger the number of generations, more accurate is the minimum number of colours required to colour the graph.\cite{geneticgc}
\end{enumerate}

    \subsubsection{Approach 3 -- Ant colony optimisation}
\begin{algorithm}
\begin{algorithmic}
% \State $x \gets 1$
% \State List $chromosomes \gets$ random colourised graphs
\Function{ant\_colony\_optimization}{$G$}
\While{$G$ is not $k$-colourable}
    \State $P$ $\gets$ Pheromone Matrix
    \While{iterations}
        \State generate\_ants()
        \For{$ant$ in $ants$}
            \State Place ant on a random vertex initially
            \State Colour the node
            \State Choose next node to travel to using the $P_{ij}$ and the $DSAT$ values of the nodes
            \State Assign a valid colour to the next node
        \EndFor{\textbf{end for}}
        \State pheromone\_decay()
        \State Update $P$ with the pheromone trail of elite ant
    \EndWhile
\EndWhile

\EndFunction
\end{algorithmic}
\caption{Interview scheduling -- Ant colony optimisation}
\label{a:antcolony}
\end{algorithm}

Algorithm 5 shows the Ant Colony Optimization approach.
This is another metaheuristic approach used for solving hard combinatorial optimization problems. It mimics the foraging behaviour of ants, and uses it to solve the graph colouring problem. The ants can indirectly communicate with each other through a  parameter called the \emph{pheromone matrix} $P$. The multitude of ants work independently but in parallel to achieve the \emph{construction graph} (The final solution where all the vertices are coloured). The ants traverse through their graphs and leave behind a pheromone trail. The pheromone trail of the best performing ant is considered, which will be reflected in $P$.

We define the pheromone matrix $P_{ij}$ as: 
\begin{equation}
    P_{ij} =
        \begin{cases}
          0 & \text{if an edge exists between node } i \text{ and node } j\\
          1 & \text{otherwise}\\
    \end{cases}    
\end{equation} and an adjacency matrix
\begin{equation}
    G_{ij} =
        \begin{cases}
          1 & \text{if an edge exists between node } i \text{ and node } j\\
          0 & \text{otherwise}\\
    \end{cases}    
\end{equation}

The first step is to create the colony of ants, where each ant is an object with certain parameters:
\begin{enumerate}
    \item \textit{alpha}: relative importance given to pheromones
    \item \textit{beta}: relative importance given to the heuristic value (DSAT)
    \item the node where the ant is set to be initialized, and a list of visited and unvisited nodes
    \item a dictionary to keep track of the colouring of the graph
\end{enumerate}

Each of these ants work on their own graphs, and try to colour it.

\begin{figure}[t]
\begin{center}
\begin{tikzpicture}
\node[circle, draw=black, thick](n1){$1$};
\node[circle, draw=black, thick, right=of n1](n2){$2$};
\node[circle, draw=black, thick, below=of n2](n3){$3$};
\node[circle, draw=black, thick, right=of n2](n4){$4$};
\node[circle, draw=black, thick, right=of n4](n5){$5$};
\node[circle, draw=black, thick, below=of n1](n6){$6$};
\node[circle, draw=black, thick, above=of n2](n7){$7$};

\draw[-] (n1) to node[above]{}(n2);
\draw[-] (n1) to node[above]{}(n6);
\draw[-] (n2) to node[above]{}(n3);
\draw[-] (n2) to node[left]{}(n4);
\draw[-] (n2) to node[above]{}(n7);
\draw[-] (n3) to node[above]{}(n6);
\draw[-] (n4) to node[above]{}(n5);
\end{tikzpicture}
\end{center}
\caption{Example graph: Panelists $p_1$, $p_2$, $p_3$, Candidates $c_1$, $c_2$, $c_3$, $c_4$}
\label{f:aco-example}
\end{figure}
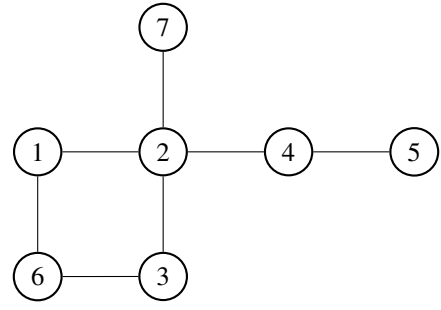

We consider an example (consider Figure~\ref{f:aco-example}) to understand how an ant colours its graph, and updates the pheromone matrix appropriately. We can take the minimum number of colours required from the Chaitin's Algorithm.

We initialize the ant on node $3$ (randomly chosen). The ant makes use of the adjacency matrix to check for its neighbours' colours, and if any of them are coloured, it makes a note of it by making it \textit{taboo} (Tabu search) \cite{tabusearch}. Once it makes note of all the taboo colours, which in our case would be none as the graph has not been coloured yet, we check for the available colours, and assign it the lowest possible colour value (0 in this case). So, after a single iteration, our graph would look like:
\begin{center}
    \textit{\{1: None, 2: None, 3: 0, 4: None, 5: None, 6: None, 7: None\}}
\end{center}

After the initial colouring, the ant has to decide where it should move. This decision is aided by $P_{ij}$ and the DSAT value. The ant now iterates through its list of unvisited nodes, and calculates the heuristic values for all of the nodes.
The heuristic value $\nu$ is defined as
\begin{equation}
    \nu = P_{ij}^{\alpha} \times dsat^{\beta}
\end{equation}

 The $P_{ij}$ value is taken from the pheromone matrix corresponding to the node the ant is residing in and the node it would visit. The DSAT value (based on the \textit{desaturation} value of all nodes in a graph \cite{dsaturation_gc}) is defined as the number of different colours among the neighbours of the node. $\alpha$ and $\beta$ here control whether we give more importance to Pheromone matrix value or the DSAT value. After we experimented with a range of values for $\alpha$ and $\beta$, it came out that prioritizing the DSAT value gives us better results, as it reduces the randomness of the pheromone matrix in the initial stages. So a higher $\beta$ over $\alpha$ would be preferred. \\
 
 The DSAT values for the unvisited nodes comes out as 
 \begin{center}
     \textit{\{1: 1, 2: 2, 4: 1, 5: 1, 6: 2, 7: 1\}}
 \end{center}(Note that \textit{None} is also counted as a colour). The $P$ value for the unvisited nodes comes out as
 \begin{center}
     \textit{\{1: 1, 2: 0, 4: 1, 5: 1, 6: 0, 7: 1\}}
 \end{center} And the heuristic values according to the formula comes out as
 \begin{center}
     \textit{\{1: 1, 2: 0, 4: 1, 5: 1, 6: 0, 7: 1\}}
 \end{center} The maximum heuristic value is taken as the ant's next destination, and in our case it could be any of $1, 4, 5$ and $7$. We assume that the ant moves to $1$. Here, we repeat the process of finding taboo colours for this particular node and colour it with the least valued colour, which would again be $0$. After this iteration the graph would look like
 \begin{center}
    \textit{\{1: 0, 2: None, 3: 0, 4: None, 5: None, 6: None, 7: None\}}
 \end{center}
 
This above process is repeated for all the unvisited nodes by the ant. After it visits all the nodes, the graph colouring would look like 
\begin{center}
    \textit{\{1: 0, 2: 1, 3: 0, 4: 0, 5: 1, 6: 1, 7: 0\}}
\end{center}

Similarly, all the other ants work on their respective graphs and colour them. After all the ants finish colouring their respective graphs, an evaluation of which ant performed the best would be conducted, the factor being the number of colours used to create the construction graph. The ant which performed the best is known as the \textit{elite ant}, and the pheromone matrix is updated based on the path the ant took to colour it, indicating to the next iteration of ants to follow it's trail to get to achieve an optimal solution. A pheromone decay is performed before the elite ant's path is updated in $P$ to even out the repetitiveness of a particular construction graph, which would be caused by increasing certain pheromone values (the path taken by elite ant). 
 
This entire process is performed for a certain number of iterations. If the dataset is small (around $100$ nodes), $2-3$ iterations would be sufficient to achieve an optimal solution.

\section{Experiments} \label{s:exp}

\subsection{Setup}
\subsubsection{Implementation}
We have implemented our interview panel creation algorithm and interview scheduling algorithm as part of a research application processing (RAP) system\footnote{\href{https://github.com/sujitkc/rap}{https://github.com/sujitkc/rap}}.

\subsubsection{Data set}
We have applied our method on two real data sets corresponding to research admissions in our institute corresponding to two different admission cycles. These two data sets are referred to as \emph{Data set 1} and \emph{Data set 2} in our results. Manual panel creation data for data set 1 was unavailable. For operational reasons, we could not obtain the (manual) panel creation and interview scheduling data for data set 1. Both these data were available for data set 2.

\subsubsection{Methodology.} We have used our algorithm to generate interview panels automatically alongside the manual process which is currently followed in our institute. We have compared the performance of our algorithm against the manual approach against multiple measures. The measures used for comparison are explained in sections ~\ref{s:pq} and ~\ref{s:sq}. The research questions we try to answer through our experiments are as follows:
\begin{enumerate}
\item $RQ_1$. How does automated panel creation compare with manual panel creation in terms of the panel quality?
\item $RQ_2$. How do the two approaches of automated panel creation -- \emph{edge ordering} and \emph{max flow} -- compare with each other in terms of panel quality?
\item $RQ_3$. How does automated scheduling compare with manual interview scheduling in terms of schedule quality?
\item $RQ_4$. How do the various approaches of automated interview scheduling -- \emph{Chaitin}, \emph{genetic algorithm} and \emph{ant colony optimisation} -- compare with each other in terms of the schedule quality?
\item $RQ_5$. Is there a difference in the schedule quality generated using the panels generated using the manual process and automated process?
\end{enumerate}

$RQ_1$ and $RQ_2$ refer to the panel generation step. $RQ_3$ and $RQ_4$ are about the interview scheduling step. $RQ_5$ is about the influence of panel generation method on the efficacy of interview scheduling. % The workflow followed in conducting the experiments is schematically shown in the block digram of Figure~\ref{f:exp}.

\subsection{Review/Interview Panel Generation}

\subsubsection{Manual Generation}
It is difficult to precisely measure the effort involved in manual generation of interview panels. However, here is our best effort estimate: Manual process of panel generation requires a several professors to sift through all of the applications. Each application is a dossier of several documents like transcripts, statement of purpose, CV and certificates. On an average each application counts up to about 35 pages. A very cursory glance would take at least 15 minutes. So, for nearly 500 applications (a typical number in our institute currently), just the sifting process takes approximate 7500 minutes or 125 hours, which is approximately 16 person days of work.

\subsubsection{Automated Generation}
In comparison, the algorithm generates the panels nearly instantaneously. The real cost of this approach is in generating the input data which the algorithm uses. Each candidate has to fill a survey comprising of a multiple choice question asking them to click on the options corresponding to their research interests. Each candidate would typically spend about 10 minutes to do this. So, 500 candidates take about 5000 minutes (approximately 2500 minutes for 500 candidates). The time thus saved may seem modest. But this is very likely deceptive. Note that:

\begin{enumerate}
\item The task of responding to the multiple-choice question is cognitively far less taxing than mapping research interests of a candidate to a group of professors.
\item The data thus generated is likely to be of much better quality, as the candidate will be able to speak of her research interests accurately. Contrary to this, an overworked professor will often superimpose her biases and ignorance in mapping an application to a group of faculty members by simply glancing through an application.
\item Since this is akin to a crowdsourced activity, the cost gets distributed to willing agents.
\item The research interests of individuals involved is currently found out through a survey. However, several documents submitted as a part of the online application are a rich source of information about the academic interests of the candidates. This information can be extracted using natural language processing~\cite{fsnlp}. Once automated, this component of manual effort also goes away making our method completely automated. We are currently working on developing this module for our RAP system.
\end{enumerate}

To summarise, in our estimate, the algorithm takes far less time in effect compared to the manual method of panel generation. Further, the data generated through the survey is of a higher quality than what a professor can manage by glancing through an application; and this fact reflects in the improved quality of the panels as discussed in the next subsection.

\subsection{Panel Quality} \label{s:pq}
We measure the quality of the panels by how well the panel use the time of the professors and the candidates. Having panelists who are not experts in the area of interest of the candidate interview/review the candidate/application affects the quality of the panel/review. Likewise, from a panelist's point of view, being part of a review panel for an irrelevant application is a waste of time and counts as negative.

\subsubsection{Candidate's Panel Quality  ($Q_C^e$)}
Thus the sum of weights of the edges connecting a candidate with her interviewer is a good measure of the panel quality from the candidate's point of view. For a candidate $c$, the panel quality can be calculated as:

\begin{equation}
Q_C^e(c) = W_I
\end{equation}

where $W_I$ is the sum of the edge weights connecting the candidate with the panelists. Superscript $e = \{R, I\}$ where $R$ stands for \emph{review} and $I$ stands for \emph{interview}. That is, we compute the quality of panels during both the review process and the interview process. For example $Q_c^R$ is the panel quality during the review process. The average candidate's panel quality is:

\begin{equation}
\bar{Q}_C^e = \frac{\sum_{c\in C^e} Q_C^e(c)}{|C^e|}
\end{equation}

where $C^e$ is the set of candidates being reviewed or interviewed.

\subsubsection{Panelist's Panel Quality ($Q_P^e$)} An interviewer's time gets used well when she interviews a candidate when there is a large overlap between her topics of interest and the candidates'. This gives us the second measure of panel quality: \emph{interviewer's Panel Quality}. For an panelist $p$, panel quality is the ratio between the edge weights connecting her to the candidates she has interviewed ($C_p^e$) and the number of candidates she has interviewed.

\begin{equation}
Q_P^e(p) = \frac{\sum_{c\in C_p^e} W(p, c)}{|C_p^e|}
\end{equation}

The average score of panelist's panel quality is:

\begin{equation}
\bar{Q}_P^e = \frac{\sum_{i\in P} Q_P^e(i)}{|P|}
\end{equation}

where $P$ is the set of all panelists.q1

We have computed the panel quality ($Q_C^e$ and $Q_P^e$) of both the manually generated panels and automatically generated panels. Our observations for two datasets (Dataset 1 and Dataset 2) each corresponding to the admission cycles of two distinct academic terms has been shown in Table~\ref{t:pq}.

We also calculated the time wastage from the panelists' perspective, by counting the number of panelists who were in panels where they could not contribute meaningfully as there was no overlap between the academic interests of the panelist with the corresponding candidates. Our observations in this regard are tabulated in Table~\ref{t:tw}. As can be seen, in the manually generated panels, panelists end up wasting more time than in automatically generated panels.

\begin{table}
\begin{center}
\begin{tabular}{|c | l| c| c|}
\hline
Panel Quality & Approach & Dataset1 & Dataset2 \\
 \hline
\multirow{3}{*}{$Q_C^R$}  & Manual &  & -0.1072503\\
  \cline{2-4}
  & Edge sorting & 0.0051055 &  0.0134299\\
  \cline{2-4}
  & Max flow & 0.0868847 &  0.0973861\\
\hline
\multirow{3}{*}{$Q_C^I$}  & Manual &  &  -0.0367124\\
  \cline{2-4}
  & Edge sorting &  &  0.0766731\\
  \cline{2-4}
  & Max flow &  &  0.2709359\\
\hline
\multirow{3}{*}{$Q_P^R$}  & Manual &  &  -1.7691448\\
  \cline{2-4}
  & Edge sorting & 1.4320453 &  1.4811494\\
  \cline{2-4}
  & Max flow & 2.4218030 &  1.4279264\\
\hline
\multirow{3}{*}{$Q_P^I$}  & Manual &  &  -0.4768062\\
  \cline{2-4}
  & Edge sorting &  &  4.2805435\\
  \cline{2-4}
  & Max flow &  &  4.2067129\\
\hline

\end{tabular}
\end{center}
\caption{Panel Quality}
\label{t:pq}
\end{table}

\begin{table}[b]
\begin{center}
\begin{tabular}{| l| c| c|}
\hline
Approach & Review Panels & Interview Panels \\
\hline
Manual & 211  &  36\\
\hline
Edge sorting & 7 & 5 \\
\hline
Max flow & 10 & 5\\
\hline

\end{tabular}
\end{center}
\caption{Time wastage}
\label{t:tw}
\end{table}

\subsection{Schedule Quality} \label{s:sq}
Schedule quality is determined as a function of elapse-time of interviews: The longer it takes to complete the interviews, the worse is the quality of the schedule. It is assumed that schedules created are feasible (i.e. conflicting interviews must not be scheduled in parallel sessions). For our research application data, we have measured the elapse time of the schedules generated by all our heuristic algorithms and compared them with the schedule generated manually and with one another. Elapse time is measured as the total number of time slots (each potentially running multiple panels) required to complete all the interviews.

\begin{table}[t]
\begin{center}
\scalebox{0.8}{
\begin{tabular}{| l | c | c |}
\hline
 Panel Generation & Interview Scheduling & Dataset2 \\
\hline
\multirow{4}{*}{Manual} & Manual & 43 \\
\cline{2-3}
 & Chaitin & 29\\
\cline{2-3}
 & GE & 29\\
\cline{2-3}
 & ACO & 29\\
\hline
\end{tabular}
}

(a)

\scalebox{0.8}{
\begin{tabular}{| l | c | c | c |}
\hline
 Panel Generation & Interview Scheduling & Dataset1 & Dataset2 \\
\hline
\multirow{3}{*}{Automated (Edge Sorting)} & Manual & \cellcolor{Gray}$\times$ & \cellcolor{Gray}$\times$ \\
\cline{2-4}
 & Chaitin & 41 & 16\\
\cline{2-4}
 & GE & 41 & 16\\
\cline{2-4}
 & ACO & 41 & 16\\
\hline
\multirow{3}{*}{Automated (Max Flow)} & Manual & \cellcolor{Gray}$\times$ & \cellcolor{Gray}$\times$ \\
\cline{2-4}
 & Chaitin & 41 & 17\\
\cline{2-4}
 & GE & 41 & 17\\
\cline{2-4}
 & ACO & 41 & 17\\
\hline
\end{tabular}
}

(b)
\end{center}
\caption{Interview Schedule Quality: (a) for manually generated panels from dataset 2 (we could not obtain the panel creation data for data set 1); (b) for automatically generated panels from dataset 1 and dataset 2.}
\label{t:sq}
\end{table}

\subsection{Experiment Workflow}
For addressing $RQ_1$ and $RQ_2$, we used the manual, edge ordering and max flow methods to generate panels for Data set 2. As mentioned above, the manual panel creation and interview scheduling data for data set 1 was not available. To address $RQ_2$ and $RQ_3$, we measured the quality of the interview schedules generated manually for Data set 2 and compared this with the quality of the interview schedules generated by all the three automated interview schedule generation approaches. The input to the automated interview scheduling algorithms were the panels as generated from the edge ordering approach. To answer $RQ_4$, we compared the results of automated approaches of interview scheduling for both the data sets. Finally, to address $RQ_5$, we fed the outputs of the manually generated panels and automatically generated panels (by edge ordering approach) to ant colony optimisation approach and compared the interview schedule qualities generated.

\subsection{Observations}

We compare the performance of our algorithms against the manual approach against multiple measures. To answer the questions raised in Section \ref{s:exp}:
\begin{enumerate}
    
    % How  does  automated  panel  creation  comparewith manual panel creation in terms of the panel qual-ity?
    \item $RQ_1$: From the results of the experiment, we can see that unsurprisingly, automated panel creation outperformed the manual panel creation in terms of panel quality from both candidate's perspective as well as panelist's perspective. The time wastage also significantly reduced in automated panel creation.\\
    
    One of the major reasons behind this is that, when panels are created manually, there is scope for an imbalanced panel. The factors like popularity of a topic among the candidate applications, complete knowledge about faculty's profile, etc. are often missed at time of manual panel creation, which causes the imbalance.
    
    % How do the two approaches of automated panelcreation –edge orderingandmax flow– compare witheach other in terms of panel quality?
    \item $RQ_2$: As we saw earlier, edge sorting approach is a non deterministic approach where as max flow approach is a deterministic approach and always guarantees an optimal solution. From the experimental results, we can observe that in terms of panel quality from candidate's perspective (which is almost same as the total cost of assignment), max flow approach outperformed the edge sorting approach. However, in terms of panel quality from panelist's perspective and time wastage, results are nearly similar for both approaches.
  
  % How a alsoand l resultsdoes an optimalutomated scheduling compare with manual interview scheduling in terms of schedule quality?
    \item $RQ_3$: We determine the schedule quality by the number of slots computed for a particular dataset. Manual scheduling is often done by a designated group whose main aim is to not cause any faculty conflicts appearing in multiple interviews. So, naturally they are not optimized to fit in a minimum number of slots, and the slots are scattered over a longer time. Table \ref{t:sq}  further reinforces the optimization of the scheduling process, as we see a drastic change in the number of slots from 43 to 16. 
    
    % How  do  the  various  approaches  of  automated interview scheduling –Chaitin,genetic algorithm and ant colony optimisation– compare with each other in terms of the schedule quality?
    \item $RQ_4$: We observe that the number of slots generated remained the same across the three scheduling algorithms that we used. The uniformity in the number of slots across the three scheduling algorithms for a particular set of generated panels can be attributed to the small size of the data sets that we used. The sole differentiating factor between the algorithms was the run time, where Algorithm \ref{a:chaitin} outperformed Algorithm \ref{a:genetic} and Algorithm \ref{a:antcolony} significantly ($\approx 10\ times$). However, it should be noted that if the number of iterations of Algorithm \ref{a:antcolony} and number of generations in Algorithm \ref{a:genetic} is low, it might lead to inefficient results.
    
    %Is there a difference in the schedule quality gen-erated using the panels generated using the manual pro-cess and automated process?%
    \item $RQ_5$: It can be seen from Table \ref{t:sq} that the results of automatically generating interview schedules from automatically generated panels is markedly superior (17 in Table~\ref{t:sq}(b)) than that when the panels are created manually (29 in Table~\ref{t:sq}(a)).
    
\end{enumerate}

\section{Related Work} \label{s:rw}

\subsection{Panel Creation}
The problem of application processing and automating it has also been done previously by some researchers in University of Michigan in their research work \cite{matching_graduate}. 
They have discussed this problem in context of the research applications that are received by U.S. universities in large number for admissions. Specifically, they worked on the step of matching research applicants with the faculty in universities using the concept of natural language processing.\\

Similar work has also been done in IBM India Research Lab \cite{ibm_prospect}. They designed a system called “PROSPECT” to screen candidates for recruitment. This system helps in automating the process of candidate-screening by automating the decision making.\\

These works provide good solution for screening the applicants and finding common areas of interests of these applicants with the potential advisors which is the initial phase of application processing.\\

Given the large number of applications, it is not possible for a person to review or interview all applications even with common areas of interest. There is need to divide the work which is why panels are created to perform this task efficiently. However, we did not find any work that specifically solves this part of application processing.\\

Hence, we felt that this process of creating the panels, which is equally important step in application processing, also requires automation.

\subsection{Interview Scheduling}
We have tackled the problem of interview scheduling by reducing it to the graph colouring problem (GCP). There have been several attempts by researchers to solve the GCP efficiently. Given that GCP is NP-complete \cite{chaitin_np_comp}, we have made use of three algorithms (\emph{Chaitin's algorithm} \cite{regallocgraphcoloring}, \emph{genetic algorithm} \cite{geneticgc}, \emph{ant colony optimization} \cite{aco_gc}) to efficiently create a schedule with zero conflicts. Chaitin's algorithm extends the traditional register allocation problem into graph colouring \cite{registeralloc_gc}. Previous research on schedule evaluation has also been done using various other meta-heuristics, one of which is the \emph{simulated annealing method} \cite{scheduling_simulated_annealing}. More research has been done on different meta-heuristics for graph colouring (chaotic ant swarm \cite{gc_chaotic_ant_swarm}, simulated anealing \cite{simulated_annealing_company_scheduling}, quantum heuristics \cite{quantumheuristics}, dynamic graph colouring \cite{dynamic_graph_coloring}, and a time-efficient demon algorithm \cite{demonalgo}).

%https://www.ncbi.nlm.nih.gov/pmc/articles/PMC6756213/
%https://pubsonline.informs.org/doi/abs/10.1287/inte.11.5.57
% https://iajit.org/PDF/vol.5,no.1/11-68.pdf

Researchers have been working on the scheduling problem for quite a while, as this paper \cite{gc_large_scheduling} written in 1979 showcases an algorithm known as RLF which works on solving large scheduling problems. However, the algorithm exhibits higher order time complexity($\mathcal{O}(n^{3})$) for certain cases when the graphs have a high edge density. And the 1981 paper \cite{gc_application_exam} by Mehta introduces the application of graph colouring to exam/course scheduling, where they talk about how economical and easy it is to equate the scheduling problem to graph colouring, and also to schedule courses parallelly.

Further research in the field of graph colouring for scheduling led to several papers such as 
\cite{gc_course_scheduling} which uses a greedy approach to colour the graph to solve their problem of course scheduling and minimize the number of conflicts. \cite{class_scheduling} applies a meta-heuristic approach known as \emph{vertex colouring} to solve their class scheduling problem. But the constraint in these papers as well as the others on exam/course scheduling is that they have a fixed number of slots to work with, so they will have to make do with personnel/time conflicts. There has not been much progress made in the interview scheduling domain, where the number of slots are not fixed, but zero conflicts in personnel is given a high priority which we have explored in this paper. With online interviews attaining more mainstream acceptance, the idea of a limited number of available rooms also loses relevance opening doors towards more aggressive parallel scheduling of interviews.

Standalone research has been done on the \emph{ant colony optimization} method to model insects and other natural phenomena \cite{antcolonyoptimization}. There has also been further research done on using a modified variant of the ant colony optimization meta-heuristic for graph colouring (modified ant colony system for colouring graphs \cite{modifiedaco_gc}). 

\section{Conclusion} \label{s:conc}
Application processing is an important and resource-intensive problem that needs to be solved by a wide variety of organisations. The current followed in practice is predominantly manual. This leads to high expense and low efficiency and effectiveness. In this paper, we have built a case in favour of automating application processing. We have presented a mathematical model of the application processing problem. This comprises of two steps: \emph{panel creation} (which maps to the assignment problem) and \emph{interview scheduling} (which maps to the graph colouring problem). We presented two algorithms for solving panel creation: \emph{edge sorting} and \emph{minimum capacity maximum flow algorithm}. We have presented three algorithms for solving interview scheduling: \emph{Chaitin's algorithm}, \emph{genetic algorithm} and \emph{ant colony optimisation}. Of these five algorithms, edge sorting is designed by us, while all the others are adapted from well known algorithms of the same name. We have evaluated the effectiveness of our approach by applying it to real data sets. For our measurement, we have defined metrics like panel quality and schedule quality. Our experiments clearly show that automation leads to significant saving of time for the selectors, and increases the effectiveness and efficiency of application processing. We position our work not as an algorithms paper, but as one presenting a systematic approach towards solving an industrial problem using mathematical modelling, discovery of new algorithms or adaptation of existing ones to solve the problem and experimental evaluation of solution approaches based on novel problem specific metrics.

In the current stage, the assignment graph is prepared through an online data collection through participation of selectors and candidates wherein they provide information about their areas of research. We believe that this information can be automatically extracted: in case of selectors -- from the organisational data; and in case of candidates -- from the wealth of documents submitted by them as a part of their application. Our future work will focus on preparing the assignment graph automatically using NLP and knowledge representation techniques.

%%References section
\bibliography{ref}

\begin{thebibliography}{10}

\bibitem{network_flows}
Ravindra~K. Ahuja, Thomas~L. Magnanti, and James~B. Orlin.
\newblock {\em Network Flows: Theory, Algorithms, and Applications}.
\newblock Prentice-Hall, Inc., USA, 1993.

\bibitem{mincostmaxflow_algo}
Zvi Galil and {\'E}va Tardos.
\newblock An o (n2 (m+ n log n) log n) min-cost flow algorithm.
\newblock {\em Journal of the ACM (JACM)}, 35(2):374--386, 1988.

\bibitem{geneticgc}
Musa Hindi and Roman Yampolskiy.
\newblock Genetic algorithm applied to the graph coloring problem.
\newblock {\em Midwest Artificial Intelligence and Cognitive Science
  Conference}, page~60, 01 2012.

\bibitem{tabusearch}
A.~Hertz and D.~de~Werra.
\newblock Using tabu search techniques for graph coloring.
\newblock {\em Computing, Springer}, 39:345--351, 1987.

\bibitem{dsaturation_gc}
Michael A.~Trick Anuj~Mehrotra.
\newblock A column generation approach for graph coloring.
\newblock {\em INFORMS Journal on Computing}, 8(4):344--354, January 1996.

\bibitem{fsnlp}
Christopher~D. Manning and Hinrich Sch\"{u}tze.
\newblock {\em Foundations of Statistical Natural Language Processing}.
\newblock MIT Press, Cambridge, MA, USA, 1999.

\bibitem{matching_graduate}
Shibamouli Lahiri, Carmen Banea, and Rada Mihalcea.
\newblock Matching graduate applicants with faculty members.
\newblock In {\em International Conference on Social Informatics}, pages
  41--55. Springer, 2017.

\bibitem{ibm_prospect}
Amit Singh, Rose Catherine, Karthik Visweswariah, Vijil Chenthamarakshan, and
  Nandakishore Kambhatla.
\newblock {PROSPECT:} a system for screening candidates for recruitment.
\newblock In {\em Proceedings of the 19th {ACM} Conference on Information and
  Knowledge Management, {CIKM} 2010, Toronto, Ontario, Canada, October 26-30,
  2010}, pages 659--668, 2010.

\bibitem{chaitin_np_comp}
Fabrice~Rastello. Florent~Bouchez, Alain~Darte.
\newblock Register allocation : what does chaitin’s npcompleteness proof
  really prove ?
\newblock {\em Laboratoire de l’informatique du parallélisme.}, pages 2--12,
  2006.

\bibitem{regallocgraphcoloring}
Gregory~J Chaitin.
\newblock Register allocation \& spilling via graph coloring.
\newblock {\em ACM Sigplan Notices}, 17(6):98--101, 1982.

\bibitem{aco_gc}
Karthikeyan R., Dr.~T. Geetha, Thamaraiselvan .C, and Sushil Kumar.
\newblock Ant colony system for graph coloring problem.
\newblock {\em International Journal of Engineering Science and Computing},
  7(7):14120--14125, July 2017.

\bibitem{registeralloc_gc}
Shengning Wu and Sikun Li.
\newblock Extending traditional graph-coloring register allocation exploiting
  meta-heuristics for embedded systems.
\newblock In {\em Third International Conference on Natural Computation (ICNC
  2007)}, volume~4, pages 324--329, 2007.

\bibitem{scheduling_simulated_annealing}
Alex Cave, Saeid Nahavandi, and Abbas Kouzani.
\newblock Schedule evaluation: Simulation optimization for process scheduling
  through simulated annealing.
\newblock In {\em Proceedings of the 34th Conference on Winter Simulation:
  Exploring New Frontiers}, WSC '02, page 1909–1913. Winter Simulation
  Conference, 2002.

\bibitem{gc_chaotic_ant_swarm}
Fangzhen Ge, Zhen Wei, Yiming Tian, and Zhenjin Huang.
\newblock Chaotic ant swarm for graph coloring.
\newblock In {\em 2010 IEEE International Conference on Intelligent Computing
  and Intelligent Systems}, volume~1, pages 512--516, 2010.

\bibitem{simulated_annealing_company_scheduling}
Xijun Lin, Qiang Lin, and Yanwei Shang.
\newblock A scheduling optimization algorithm based on graph theory and
  simulated annealing.
\newblock In {\em 2021 6th International Conference on Inventive Computation
  Technologies (ICICT)}, pages 492--496, 2021.

\bibitem{quantumheuristics}
Alex Fabrikant and Tad Hogg.
\newblock Graph coloring with quantum heuristics.
\newblock In {\em AAAI/IAAI}, pages 22--27, 2002.

\bibitem{dynamic_graph_coloring}
Long Yuan, Lu~Qin, Xuemin Lin, Lijun Chang, and Wenjie Zhang.
\newblock Effective and efficient dynamic graph coloring.
\newblock {\em Proc. VLDB Endow.}, 11(3):338–351, nov 2017.

\bibitem{demonalgo}
Amani~A. Alahmadi, Taghreed~M. Alamri, and Manar~I. Hosny.
\newblock Time efficient demon algorithm for graph coloring with search cut-off
  property.
\newblock In {\em 2014 Science and Information Conference}, pages 254--259,
  2014.

\bibitem{gc_large_scheduling}
Leighton.
\newblock A graph coloring algorithm for large scheduling problems.
\newblock {\em J Res Natl Bur Stand (1977)}, 84(6):489–506, November 1979.

\bibitem{gc_application_exam}
Nirbhay~K. Mehta.
\newblock The application of a graph coloring method to an examination
  scheduling problem.
\newblock {\em INFORMS Journal on Applied Analytics}, 11(5):57--65, October
  1981.

\bibitem{gc_course_scheduling}
Ammar Elhassan.
\newblock Graph-coloring for course scheduling — a comparative analysis based
  on course selection order.
\newblock In {\em The Third International Conference on e-Technologies and
  Networks for Development (ICeND2014)}, pages 83--88, 2014.

\bibitem{class_scheduling}
Amal Dandashi and Mayez Al-Mouhamed.
\newblock Graph coloring for class scheduling.
\newblock In {\em ACS/IEEE International Conference on Computer Systems and
  Applications - AICCSA 2010}, pages 1--4, 2010.

\bibitem{antcolonyoptimization}
Marco Dorigo, Mauro Birattari, and Thomas Stutzle.
\newblock Ant colony optimization.
\newblock {\em IEEE Computational Intelligence Magazine}, 1(4):28--39, 2006.

\bibitem{modifiedaco_gc}
TaeChoong~Chung SangHyuck~Ahn, SeungGwan~Lee.
\newblock Modified ant colony system for coloring graphs.
\newblock In {\em Fourth International Conference on Information,
  Communications and Signal Processing, 2003 and the Fourth Pacific Rim
  Conference on Multimedia. Proceedings of the 2003 Joint}, volume~3, pages
  1849--1853 vol.3, 2003.

\end{thebibliography}

%\begin{thebibliography}{99} 
%\bibitem{latexcompanion} 
%Michel Goossens, Frank Mittelbach, and Alexander Samarin. 
%\newblock {\em The \LaTeX\ Companion}. 
%Addison-Wesley, Reading, Massachusetts, 1993.
%\end{thebibliography}

\end{document}